# Tunable bifurcation of magnetic anisotropy and bi-oriented antiferromagnetic order in kagome metal GdTi$_3$Bi$_4$

Jianfeng Guo[1,*], Shiyu Zhu[1,2,*,†], Runnong Zhou[1,2,*], Ruwen Wang[1,2,*], Yunhao Wang[1,2], Jianping Sun[1,2], Zhen Zhao[1,2], Xiaoli Dong[1,2], Jinguang Cheng[1,2], Haitao Yang[1,2,‡], Jiang Xiao[3,§], and Hong-Jun Gao[1,2,∥]

[1]*Beijing National Center for Condensed Matter Physics and Institute of Physics, Chinese Academy of Sciences, Beijing 100190, China.*
[2]*School of Physical Sciences, University of Chinese Academy of Sciences, Beijing 100190, China.*
[3]*Department of Physics and State Key Laboratory of Surface Physics, Fudan University, Shanghai 200433, China.*

The novel kagome family RTi$_3$Bi$_4$ (R: rare-earth) offers a unique platform for exploring distinctive physical phenomena such as anisotropy, spin density wave, and anomalous Hall effect. In particular, the magnetic frustration and behavior of magnetic anisotropy in antiferromagnetic (AFM) kagome materials are of great interest for the fundamental studies and hold promise for next-generation device applications. Here, we report a tunable bifurcation of magnetic anisotropic and bi-oriented AFM order observed in the quasi-1D kagome antiferromagnet GdTi$_3$Bi$_4$. The magnetic domain evolutions during two plateau transition processes are directly visualized, unveiling a pronounced in-plane anisotropy along the a-axis. Temperature-dependent characterization reveals a bifurcation transition of anisotropy at approximately 2 K, where the a-axis anisotropy splits into two special orientations, revealing a hidden bi-oriented in-plane AFM order deviating from the high-symmetry direction by ±7°. More intriguingly, the characteristics of the bifurcated anisotropy are clearly illustrated through vector magnetic field modulation, revealing three distinct in-plane domain phases in the transverse magnetic field phase diagram. Our results not only provide valuable insights into the tunable bifurcation of magnetic anisotropic in GdTi$_3$Bi$_4$, but also pave a novel pathway for AFM spintronics development.

Kagome lattice materials have emerged as a forefront of research in condensed matter physics due to the presence of frustration, correlation, and non-trivial band topology [1–5]. Kagome magnets, being an important classification within this category, exhibit plenty of fascinating physical phenomena such as Chern and Weyl topological magnetism [6–9], anomalous Hall effect (AHE) [10–12], and charge density wave [13,14]. The kagome lattice materials with an antiferromagnetic (AFM) ground state are particularly interesting due to the interplay between the AFM order and the unique electronic structure of the kagome lattice [15–18]. Therefore, the investigation and exploration of physical properties inherent in novel AFM kagome materials have emerged as a prominent research focus within the field of kagome magnets.

The novel layered GdTi$_3$Bi$_4$, a member of the RTi$_3$Bi$_4$ family (R = rare-earth metals), is one of such AFM Kagome materials characterized by high anisotropy due to its unique 1D AFM zigzag (ZZ) chain structure along the a-axis. It also exhibits a variety of exotic physical phenomena, such as fractional magnetization plateau, flat bands, spin density wave, AHE and topological Hall effect (THE) [19–30]. The behavior of the magnetic anisotropy of materials is often coupled with a variety of physical properties, including magnetic, electronic, and optical characteristics, thereby offering promising avenues for manipulating collective excitations such as excitons, phonons, or magnons [31,32]. While ferromagnetic (FM) systems have been extensively studied for their controllable magnetic anisotropy in multifunctional devices [33–35], antiferromagnetic (AFM) systems have distinct advantages, such as the absence of stray fields, ultrafast spin dynamics, and resistance to magnetic perturbations, making them promising for next-generation device applications [36–38].

Here, we report a tunable bifurcation of magnetic anisotropy axis and the concomitant formation of bi-oriented in-plane AFM domains in the quasi-1D kagome metal GdTi$_3$Bi$_4$ single crystals. By employing ultra-low-temperature magnetic force microscopy (MFM) with a vector magnetic field, we directly visualize the behavior of magnetic domains in GdTi$_3$Bi$_4$ at 0.4 K during two plateau transitions: from the AFM ground state to the 1/3 magnetization plateau (first transition) and from the 1/3 magnetization plateau to the forced FM state (second transition). This reveals a pronounced planar anisotropy along the a-axis. More importantly, the temperature-dependent characterizations uncover the controllable bifurcation transition of magnetic anisotropy axis from the a-axis to two specific directions, deviating from the high-symmetry a-axis by ±7°, accompanied by the formation of bi-oriented in-plane AFM domains. The in-plane AFM order, whose Néel vector aligns with one or both of the two anisotropic axes, can be modulated using an in-plane magnetic field, illustrating the in-plane domain phase diagram comprising three distinct AFM phases. Our results reveal the tunable bifurcation of magnetic anisotropy and bi-oriented AFM order in GdTi$_3$Bi$_4$, offering a novel avenue for AFM-based programmable magnetic devices, advancing the development of next-generation logic devices.

*These authors contributed equally to this work.
†Contact author: syzhu@iphy.ac.cn (S. Z.)
‡Contact author: htyang@iphy.ac.cn (H. Y.)
§Contact author: xiaojiang@fudan.edu.cn (J. X.)
∥Contact author: hjgao@iphy.ac.cn (H.-J. G.)

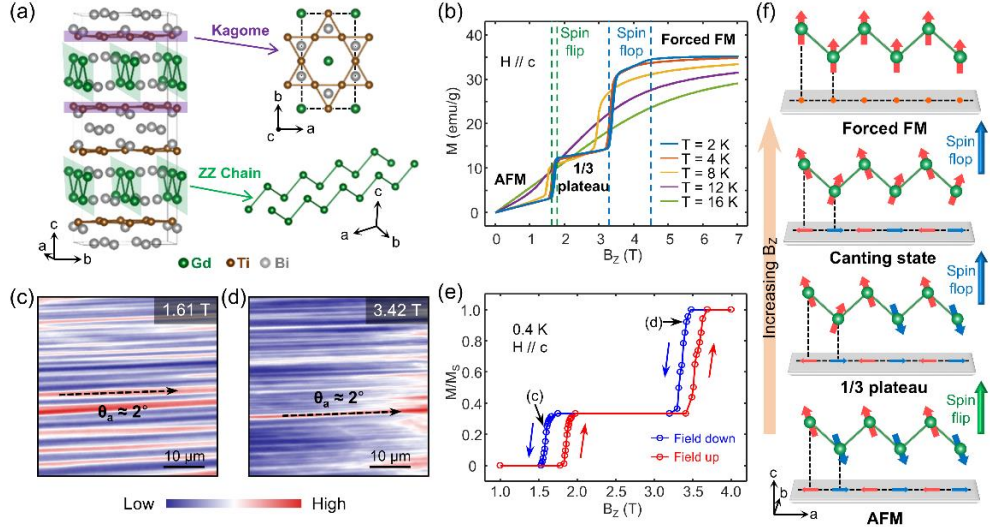

FIG. 1. Magnetic properties and anisotropic domain evolution in GdTi$_3$Bi$_4$ at 0.4 K. (a) Crystal structure of GdTi$_3$Bi$_4$ showing the Ti kagome layers and Gd ZZ chains. (b) Field-dependent magnetizations at different temperatures with H//c for GdTi$_3$Bi$_4$. (c,d) MFM images of the first (c) and second (d) transition processes in GdTi$_3$Bi$_4$ at 0.4 K. Red (blue) represents repulsion (attraction) regions. (e) M-H hysteresis loop extracted from the H-dependent MFM images. (f) Schematic illustrating the proposed spin configuration evolution with magnetic field.

GdTi$_3$Bi$_4$ is a van der Waals (vdW) AFM kagome metal, corresponding to an orthorhombic *Fmmm* (No. 69) space group with inversion symmetry. The Gd-based quasi-1D ZZ chains are sandwiched between two Ti-based kagome layers, forming a unique crystal structure [Fig. 1(a)]. Among them, the ZZ chains are aligned along the crystallographic a-axis with a displacement of one atomic position between neighboring chains. The high quality of the GdTi$_3$Bi$_4$ single crystals used in the experiments is verified by X-ray diffraction, energy-dispersive X-ray spectroscopy, and scanning electron microscopy. The magnetization and transport measurements reveal that GdTi$_3$Bi$_4$ exhibits an AFM ground state with a Néel temperature ($T_N$) of approximately 14.5 K [39].

Further field-dependent measurements uncover a fractional plateau at 1/3 of the saturation magnetization in GdTi$_3$Bi$_4$, highlighting the pivotal influence of frustrated magnetism [40,41] [Fig. 1(b)]. The emergence of the 1/3 plateau phase involves two plateau transitions, which are a spin-flip transition (referred to as the first transition) from the AFM ground state to the 1/3 plateau and a spin-flop transition (the second transition) from the 1/3 plateau to the forced FM phase.

To directly explore the behavior of magnetic anisotropy and frustrated magnetism in GdTi$_3$Bi$_4$ in real space, we performed MFM characterization to investigate the two out-of-plane plateau transitions, as shown in Figs. 1(c) and 1(d) [39]. In MFM images, the stripe domains in red and blue represent the repulsive and attractive forces sensed by the MFM tip, corresponding to the AFM and 1/3 plateau phases in the first transition [Fig. 1(c)], and to the 1/3 plateau and forced FM phases in the second transition [Fig. 1(d)]. Throughout the process of varying out-of-plane magnetic fields at 0.4 K, the elongation of the stripe domains shows a highly in-plane anisotropy, maintaining a consistent angle of 2° relative to the scanning x-direction, while ruling out the spin collective behavior in the frustrated magnetism. This special orientation has been validated to align with the ZZ chains (a-axis), as confirmed by the Laue diffraction pattern [39]. These observations strongly suggest a pronounced in-plane magnetic anisotropy along the a-axis of GdTi$_3$Bi$_4$ at 0.4 K, indicating that the spin consists of components in both the a- and c-axis directions during the entire spin transition process. This is in good agreement with the easy axis previously reported based on M-H results [25].

Subsequently, the normalized magnetization curve was extracted from the variable-field MFM images at 0.4 K [Fig. 1(e)], which exhibits good agreement with the M-H curves [see Fig. 1(b) @ 2 K and 4 K]. It is noteworthy that a distinct magnetic hysteresis was observed at 0.4 K, which is uncommon in AFM systems.

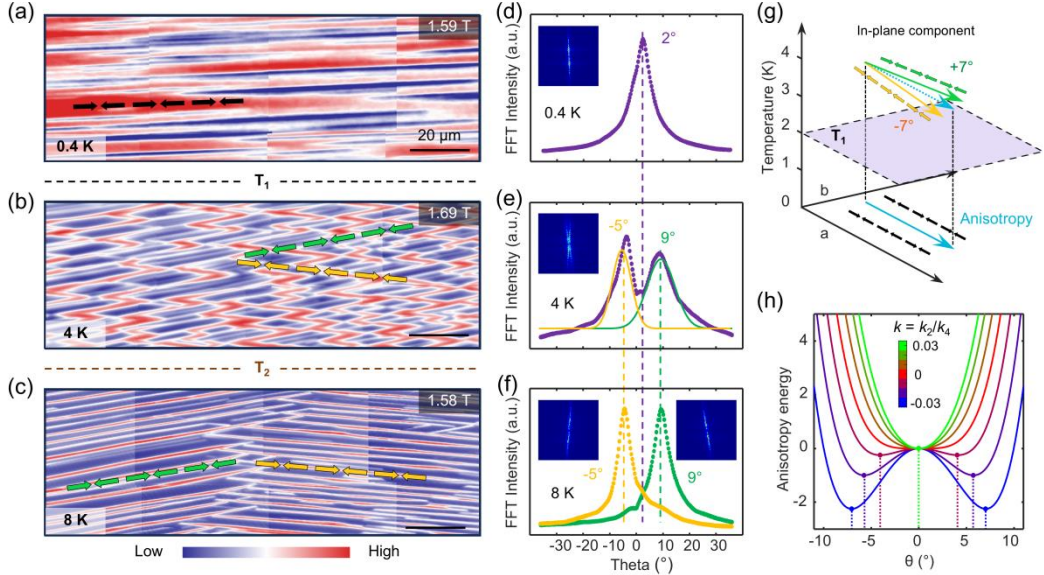

FIG. 2. Temperature-dependent bifurcation transition of magnetic anisotropy. (a-c) Typical large-scale MFM images of GdTi$_3$Bi$_4$ during the first transition at 0.4 K (a), 4 K (b) and 8 K (c). (d-f) Intensity-angle curves (scanning x-axis set to 0°) extracted from the FFT results of MFM images at 0.4 K (d), 4 K (e), and 8 K (f). Insets show the corresponding FFT images. (g) Schematic of the anisotropic bifurcation transition with temperature. (h) The plot of anisotropy energy $E_k$ as a function of in-plane direction $\theta$ for varying $k = k_2/k_4$.

This phenomenon is likely associated with the relatively strong FM coupling in the plateau and saturated phases, suggesting a uniform magnetic structure and relatively high degree of symmetry.

Based on the above observations, the evolution of the spin configurations with increasing magnetic field at 0.4 K is depicted in Fig. 1(f) [39]. In the AFM and 1/3 plateau phases at low magnetic fields, the magnetic moments exhibit components along both the a- and c-axis directions, while maintaining a consistent in-plane projection of the AFM order along the a-axis. As the field strength increases, a spin-flop transition occurs, giving rise to the canting phase. In this phase, uniform magnetic domains lead to the absence of observable contrast in MFM images. This lack of contrast persists in the subsequent forced FM phase, therefore we treat the canting phase and forced FM phase as the same phase in this manuscript, as shown in Fig. 1(e).

Next, we perform the MFM measurement at varying temperatures and observe a bifurcation of the magnetic anisotropy axis, resulting in bi-oriented anisotropy that deviates from the high-symmetry directions. With increasing temperature, GdTi$_3$Bi$_4$ display two distinct evolutionary behaviors ($T_1$ ~ 2 K and $T_2$ ~ 8 K) of the magnetic domains during the first transition, indicating a transition in anisotropy and spin configuration [Figs. 2(a)-2(c)]. For temperature near $T_1$, the elongation direction of the stripe domains gradually deviates from the a-axis direction and split into two special orientations. For temperature near $T_2$, the fragmented stripe domains in two special orientations become concentrated and merge into larger single-domain regions. We quantified the domain orientations using intensity-angle curves derived from the FFT results of representative MFM images [Figs. 2(d)-2(f)]. While a single peak is observed at 2° (a-axis) at 0.4 K, two distinct peaks emerge at -5° and 9° at 4 K, corresponding to a ±7° tilt relative to the a-axis. At 8 K, two large domains also exhibit tilting by ±7° without any further evolution.

The above temperature-dependent measurement strongly points to an in-plane anisotropic bifurcation transition above ~ 2 K, shifting the anisotropy from the a-axis to two specific directions that deviate from the a-axis by ±7°, which reorients the in-plane AFM Néel order away from the high-symmetry direction [Fig. 2(g)] [39].

Based on the Ginzburg-Landau theory, we assume the form of the anisotropy energy for the system as $E_k = k_2 \sin^2\theta + k_4 \sin^4\theta$, where $\theta$ is the angle between the in-plane magnetization and the a-axis, and $k_2$ and $k_4$ are the phenomenological parameters accounting for the uniaxial and biaxial anisotropies. The ratio $k = k_2/k_4$ critically determines the energy landscape of the system as seen in Fig. 2(h). We assume that $k_2$ and $k_4$, and consequently the ratio $k$, may vary with temperature. When $k > 0$, the minimum energy occurs at $\theta = 0°$, identifying the a-axis as the easy axis. However, when k < 0, the energy landscape exhibits bifurcation, with two minima emerging, indicating two tilted easy axes. Experimentally observed tilting angle $\theta = ± 7°$ corresponds to a ratio of $k \approx -0.03$.

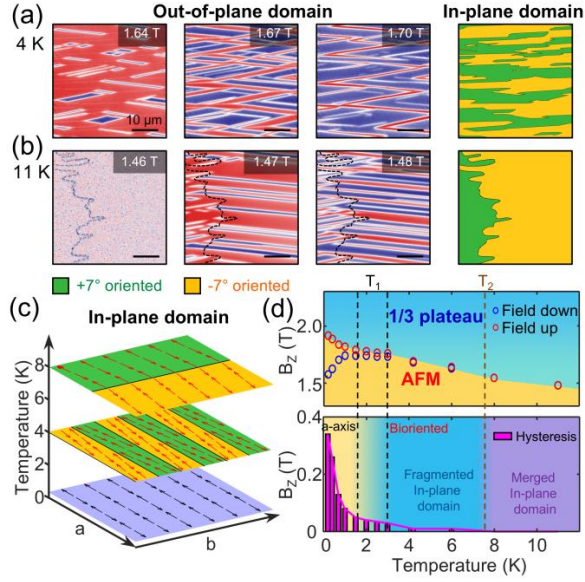

FIG. 3. In-plane oriented AFM domains. (a,b) Out-of-plane magnetic domain evolution and deduced in-plane magnetic domains at 4 K (a) and 11 K (b). (c) In-plane domain characteristics at 0.4 K, 4 K, and 8 K. (d) H-T phase diagram with magnetic hysteresis behavior of the first transition. The a-axis and bi-oriented regions are distinguished by $T_1$, while $T_2$ subdivides the bi-oriented region into fragmented and merged in-plane domains.

As a result of in-plane anisotropic bifurcation transition, the competition between the two anisotropic directions leads to the formation of fragmented in-plane AFM domains (+7° domain and -7° domain for the in-plane Néel vector, 4 K), which supposedly should be invisible in MFM images. However, these in-plane AFM domains can influence the expansion of the out-of-plane magnetic domains that is visible in MFM images. Therefore, it is possible to infer the in-plane AFM domain from the growing pattern of the out-of-plane domains. By analyzing the evolutionary paths of stripe domains with different elongation directions [39], the in-plane AFM domains at 4 K are mapped out in Fig. 3(a).

With a further increase in temperature, the fragmented in-plane domains merge into larger single domains (above 8 K), resulting in extensive regions of +7° and -7° stripe domains. Notably, we observed the magnetic signatures of the domain walls (DWs) for the in-plane AFM domains at 11 K [1.46 T, Fig. 3(b)] due to their higher magnetic susceptibility compared to other regions, which strongly validates the presence of in-plane domains. Based on the above results, we have plotted schematic diagrams illustrating the in-plane magnetic domain characteristics at three temperatures: 0.4 K, 4 K, and 8 K [Fig. 3(c)]. Moreover, it is important to emphasize that the in-plane domains exhibit robustness and are almost unaffected by an out-of-plane magnetic field. These domains respond primarily to temperature. Once formed above $T_1$, they exhibit minimal creep under varying out-of-plane magnetic fields, yet their presence remains the key factor influencing the evolution of out-of-plane magnetic domains.

The detailed variable temperature characterization reveals two distinct magnetic domain phase transitions within the temperature ranges of 1.75 K-3 K ($T_1$) and 7 K-8 K ($T_2$), as determined by extracting the orientation angles of the out-of-plane stripe domains [39]. It is worth noting that the observed gradual tilt orientations with increasing temperature align well with the phase transition model [Fig. 2(h)]. The evolution of hysteresis provides additional support for the occurrence of the bifurcation transition [Fig. 3(d)]. The hysteresis of the $GdTi_3Bi_4$ essentially diminishes with the emergence of in-plane magnetic domains, suggesting a reconfiguration of spin during the phase transition interval [42–44]. In addition to the alteration in spin configuration, the bifurcation transition of magnetic anisotropy in $GdTi_3Bi_4$ is accompanied by transformations in other physical properties, such as the recently reported unconventional charge-spin density wave [30], all of which deserve more in-depth studies in the future.

The existence of two close-by anisotropic axes offers a unique opportunity to manipulate the AFM order. To further explore the bi-oriented properties and identify the existence of in-plane domains, we conducted additional characterization using vector magnetic fields. The in-plane AFM order tends to align perpendicular to the in-plane field. The special orientation of the stripe domains is precisely controlled by adjusting the direction of the in-plane field. Depending on the elongation orientation of the magnetic domains, it is possible to partition the in-plane vector magnetic field space into three distinct regions corresponding to bi-oriented (Phase I, central area), +7° oriented (Phase II, in quadrant two and four), and -7° oriented (Phase III, in quadrant one and three) [Fig. 4(a)]. The regulation of orientation is correlated with the radial strength ($|B_{//}|$) and direction of the vector field ($\theta_B$). The switching of AFM order between different phases is demonstrated by rotating the in-plane vector field along a circular path ($|B_{//}| = 0.9$ T, $\theta_B$ ranging from 0° to 360° in 45° intervals) [Fig. 4(b)] and an arc trajectory ($|B_{//}| = 1.2$ T, $\theta_B$ ranging from 200° to 170° in 3° intervals) [39]. Notably, throughout the entire modulation of the in-plane field, only two distinct orientations at ±7° persist, further confirming the robustness of the two in-plane anisotropy axes present in $GdTi_3Bi_4$. More importantly, our experimental results validate the feasibility of achieving control over magnetic order in the AFM system through the construction of dual anisotropic axes, thereby presenting a novel operational mode for AFM devices.

With two easy axes at $\theta = \pm 7°$, distinct in-plane Néel orders are established along these axes, giving rise to in-plane AFM domains. When an in-plane magnetic field $B$ is applied, the ±7° AFM configurations experience different energy states. The Néel order inherently favors the configuration with the lower energy. If the energy difference between the two configurations is comparable to or less than the thermal energy threshold ($k_BT$), both configurations can coexist. Conversely, if the energy difference surpasses the thermal energy, only the configuration with the lower energy stabilizes, resulting in a single domain. The energy difference induced by the external field is quantified as $\Delta E \propto B^2|\sin(2\gamma)|$ with $\gamma$ the angle between B field and the easy-axis (±7° from a-axis) under consideration. The phase boundary separating the bi-stable region from the uni-stable region is given by the condition $\Delta E \sim k_BT$, which is plotted as the dashed curves in Fig. 4(a) [39], in agreement with the experimental data marked as dots.

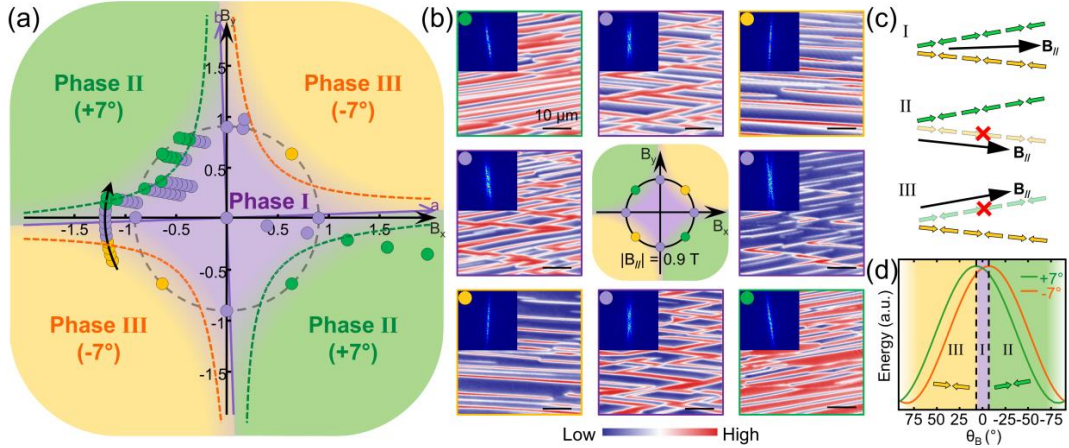

FIG. 4. Tunable oriented switching by in-plane vector magnets at 4 K. (a) $B_x$-$B_y$ phase diagram of GdTi$_3$Bi$_4$, showing three magnetic phases with different stripe domain orientations, namely bi-oriented (phase I), +7° oriented (phase II) and -7° oriented (phase III). (b) A series of typical MFM images and corresponding FFT images during continuous rotation of the in-plane vector field. (c) Schematic diagrams of tunable oriented switching process. (d) The energy comparison curves of +7° and -7° oriented domains under different in-plane magnetic field directions.

The tunable orientation switching in GdTi$_3$Bi$_4$ is fully consistent with our theoretical model, which evaluates the energy differences between the two in-plane domains under varying in-plane fields [Figs. 4(c) and 4(d)]. In contrast to conventional FM domains, the AFM domains exhibit instability and higher energy when subjected to a magnetic field aligned with their orientation. Owing to the reduced alignment between the alternative orientation and the magnetic field, a lower energy state is attained, thereby inducing a reorientation of the in-plane domain and achieving a switch in orientation. Even within a large single orientation region (8 K), the modulation of magnetic domain orientation can still be achieved [39].

Furthermore, the bifurcation transition also influences the second transition in GdTi$_3$Bi$_4$, leading to the emergence of topological spin textures and rendering the magnetic domain orientation more randomized [39]. This discovery provides a possible explanation for the observed high-resistance state and the pronounced AHE (superimposed with THE) in the 1/3 plateau phase. Finally, by integrating all the experimental data, we constructed a comprehensive $H$-$T$ phase diagram for GdTi$_3$Bi$_4$ [39].

In summary, we have performed a comprehensive real-space characterization of the magnetization plateau transition and identified a temperature-dependent in-plane anisotropic bifurcation transition in the quasi-1D kagome antiferromagnet GdTi$_3$Bi$_4$. Through the observation of the splitting of the elongation orientation of the strip domains for the out-of-plane magnetization, we reveal a bifurcation of the magnetic anisotropy axis from the high-symmetry direction in the hidden in-plane AFM order. The presence of two anisotropic axes facilitates the controlled manipulation of AFM order using an in-plane vector magnetic field, providing a new strategy for tuning the Néel order in AFM spintronic devices. Our findings not only provide a critical real-space perspective for understanding the physical phenomena in GdTi$_3$Bi$_4$, but also pave the way for novel avenues in the advancement of next-generation device applications.

*Acknowledgments*—The authors are thankful to Ziqiang Wang, Wei Ji, Cong Wang and Hengxin Tan for useful discussion. This work was funded by the National Natural Science Foundation of China (Nos. 62488201, 12374199, 12474110), the National Key Research & Development Projects of China (Nos. 2022YFA1204100), and the Beijing Nova Program (Nos. 20240484651).